\documentclass[12pt,preprintnumbers,aps,amssymb,nofootinbib,amsmath]{revtex4}
\usepackage{epsfig,epsf}
\usepackage{graphicx}
\usepackage{epstopdf}
\newcommand{\beq}{\begin{equation}}
\newcommand{\eeq}{\end{equation}}
\def\eq#1{{(\ref{#1})}}
\def\fig#1{{Fig.~\ref{#1}}}

\newcommand{\im}{\mathrm{Im}}
\begin{document} 

%\preprint{Draft \#1}

\title{Pair production by boost-invariant fields\\ in comoving coordinates}

\author{ Kirill Tuchin$\,^{a,b}$}
\affiliation{
$^a\,$Department of Physics and Astronomy, Iowa State University, Ames, IA 50011 \\
$^b\,$RIKEN BNL Research Center,
Upton, NY 11973-5000}

\date{\today}

\begin{abstract} 
We  derive the pair-production probability in a constant electric field in Rindler coordinates in a quasi-classical approximation. Our result is different from the pair-production probability in an inertial frame (Schwinger formula). In particular, it exhibits non-trivial dependence on rapidity and deviation from Gaussian behavior at small transverse momenta. 
Our results can be important for analysis of particle production in heavy-ion collisions.  

\end{abstract}

\maketitle

%%%%%%%%%%%%%%%%%%%%%%%%%%%%%%%%%%%%%%%%

\section{Introduction}
In recent years  a novel approach to particle production in high energy QCD has been developed. It is based on the observation that the color field of a high energy nucleus is a quasi-classical non-Abelian Weizsaecker-Williams field  which strength increases with energy (see reviews \cite{Iancu:2003xm,Jalilian-Marian:2005jf} and references therein). The high field strength provides a semi-hard scale at which the coupling constant is small. This makes the problem of particle production in heavy ion collisions solvable in principle using the small coupling expansion with subsequent resumation of powers in strong field. This problem was solved at the lowest order in the density of  color sources in \cite{Kovner:1995ja,Kovchegov:1997ke}. 
The resulting field is boost invariant, i.e. it depends on the invariant time $\tau=\sqrt{2x_+x_-}$ and in general, on transverse coordinate ${\bf x}_\bot$, but not on pseudo-rapidity $\eta= \frac{1}{2}\ln\frac{x_+}{x_-}$. 
Moreover, at early times after the collision $\tau\rightarrow 0$ the transverse fields vanish while the longitudinal survive \cite{Fries:2005yc}. Possible implications of this result were discussed in \cite{Lappi:2006fp}.

Existence of the strong longitudinal chromoelectric field in high energy nuclei collisions   was first pointed out in the  framework of the High Parton Density QCD in \cite{Kharzeev:2005iz}. In Ref.~\cite{Kharzeev:2006zm} it was further argued  that dependence of the produced field potential on the transverse coordinates can be neglected in the leading (transverse) logarithmic approximation implying vanishing longitudinal chromomagnetic field. 

The role of  the strong \emph{longitudinal chromoelectric field} in particle production in heavy-ion collisions was discussed in many publications  \cite{Biro:1984cf,Kajantie:1985jh,Bialas:1986mt,Kerman:1985tj,Gatoff:1987uf,Kluger:1991ib}. It was realized that the strong confining force produces particle pairs out of vacuum by means of the Schwinger mechanism \cite{Schwinger:1951nm,Casher:1978wy}.  Since at high energies and in central collisions the nuclear color field is the Weizsaecker-Williams one,  it became important to understand the pair production mechanism in the background of such field. The pair production probability was calculated in the quasi-classical approximation in  Refs.~\cite{Kharzeev:2005iz,Kharzeev:2006zm}.
It was argued in \cite{Kharzeev:2005iz} that the pair production is a mechanism by which the Color Glass Condensate can ``evaporate" into thermalized quark-gluon medium over short time hence shading some light on a problem of fast thermalization in heavy-ion collisions at RHIC. 

In the present paper we reformulate the problem of pair production in heavy ion collisions in the Rindler coordinates which are the most natural for expressing  the boost-invariance of the produced field \cite{Feynman:1973xc,Casher:1974vf,Bjorken:1982qr,Andersson:1983ia}.  In hydrodynamical models these coordinates are the coordinates of a fluid in a comoving frame. 
Since the Rindler  coordinates correspond to the rest frame of the uniformly accelerated observer, the physical processes in the Rindler frame will in general look differently than in the inertial frame. In particular, we will argue that the probability of pair production by a constant chromoelectric field  \eq{prob} is significantly different from the Schwinger result \cite{Schwinger:1951nm} in contradiction to the common belief.  This result may appear counterintuitive at first sight. However, I would like to stress that we not just use a different set of coordinates, but most importantly, quantize the system in a different space-time. 
It must be kept in mind that the Fock space in the Rindler space-time is physically different from the Fock space in the Minkowski space-time. In particular, the vacuum states $|0\rangle_R$ and $|0\rangle_M$ in the Rindler   and Minkowski space-time are different. 
This observation has been used by Unruh \cite{Unruh:1976db} to derive his famous effect: a uniformly accelerated detector (with acceleration $a$) becomes excited as if it was placed in a thermal bath at temperature $T=a/(2\pi)$. Many other examples of that kind are considered in Ref.~\cite{Birrell:1982ix}. We discuss this further in Sec.~\ref{sec:disc}.

The actual calculation of the pair production probability is carried out in the quasi-classical approximation pioneered in \cite{Brezin:1970xf,popov,Marinov:1977gq}, see \cite{Dunne:2004nc} for a review.
 
 %%%%%%%%%%%%%%%%%%%%%%%%%%%%%%%%%%%%%
\section{Pair production in the Rindler space}\label{sec:pp}

Consider a charged scalar field $\phi$ coupled to the Abelian gauge field
$A_\mu$. Equations of motion read
\begin{subequations}
\beq\label{eqmot1}
D_\mu D^\mu\phi\,+\, m^2\phi\,=0\,,
\eeq
\beq\label{eqmot2}
\partial_\mu F^{\mu\nu}\,=\,j^\nu\,.
\eeq
\end{subequations}
where $D_\mu=\partial_\mu+ieA_\mu$ and the current is
\beq\label{tok}
j_\mu\,=\,-i[\phi^*D_\mu\phi\,-\,(D_\mu\phi)^*\phi]\,.
\eeq
In absence of the back-reaction of a produced matter on the field the light-cone potentials at the leading order in density of color charge read \cite{Kovner:1995ja}
\beq\label{potent}
A_+\,=\,x_+\alpha(\tau)\,,\quad A_-\,=\,-x_-\alpha(\tau)\,.
\eeq 
where we introduced the light-cone variables $x_\pm=(t\pm x)/\sqrt{2}$; $\tau=\sqrt{2x_+x_-}$ is the invariant time.

Writing down \eq{potent} we assumed that the field does not depend on transverse coordinates. Indeed,  it was argued in \cite{Kharzeev:2006zm} that 
production of low transverse momentum particles ($p_T\ll Q_s$) is dominated by the longitudinal chromo-electric fields. This implies that particles produced by longitudinal chromo-electic fields via the pair production mechanism constitute the bulk of the particle yield \cite{com}.

In this approximation, the electric field is given by
\beq\label{ef}
E_z\,=\, -\frac{1}{\tau}\,\partial_\tau (\alpha\,\tau^2)\,.
\eeq
In this section we proceed neglecting the back-reaction  and return to this problem later on in Sec.~\ref{sec:br}.  

The Rindler coordinates  $(\eta,\tau,{\bf x_\bot})$ are simply related to the Minkowski coordinates $(t,x,{\bf x_\bot})$  by the following relationships:
\beq
\tau^2=t^2-x^2\,,\quad \eta=\frac{1}{2}\ln\frac{x_+}{x_-}\,.
\eeq
In the Rindler coordinates Eq.~\eq{eqmot1}  reads
\beq\label{phirindler}
\frac{1}{\tau}\partial_\tau(\tau\,\partial_\tau \phi)\,-\,\frac{1}{\tau^2}\partial^2_{\eta}\phi\,-\,\nabla^2_\bot\phi\, +\, 2i\, e\, \alpha\,\partial_\eta\phi\,+\, e^2\, \alpha^2\, \tau^2\, \phi\,=\,0\,.
\eeq
Switching over to the momentum space using
\beq\label{fourier}
\phi(\eta,{\bf x}_\bot, \tau)\,=\, \int \frac{d p_\eta d^2 {\bf p}_\bot }{(2\pi)^3}
e^{i p_\eta \eta-i {\bf p}_\bot\cdot {\bf x}_\bot}\, \varphi(p_\eta, {\bf p}_\bot, \tau)\frac{1}{\sqrt{\tau}}\,
\eeq
yields the following equation for $\varphi$:
\beq\label{osc}
\ddot \varphi\,+\, \omega(\tau)\varphi\,=\,0\,, \quad 
\omega^2(\tau)\,=\,\frac{1}{4\tau^2}\,+\, p_\bot^2\,+\, \bigg(
\frac{p_\eta}{\tau}\,-\, e\,\alpha\, \tau\bigg)^2\,.
\eeq 
We can express the momentum $p_\eta$ in terms of the light-cone momenta of an inertial observer using  the tensor transformation rule
\beq\label{trans}
p_\eta\,=\,\frac{\partial x_+}{\partial \eta}\,p_-\,+\, \frac{\partial x_-}{\partial \eta}\, p_+\,=\, x_+\,p_-\,-\,x_-\,p_+\,. 
\eeq
For massless particles  $p_\eta=0$. 

We are seeking solution of \eq{osc} in the form $\varphi=e^{iS}$. Introducing the auxiliary function $u=\dot S$ we have
\beq\label{ww}
u^2\,-\, i\dot u\,-\, \omega^2\,=\,0\,.
\eeq
Assuming that $S$  is slowly varying function so that $|\ddot S|\ll \dot S^2$ 
 we can adiabatically expand $u=u_0+u_1+\ldots$ where the functions $u_0$ and $u_1$ satisfy $u_0^2-\omega^2=0$ and
$2u_0u_1-i\dot u_0=0$. Thus, $u=\pm \,\omega+i\dot \omega/2\omega+\ldots$ and 
\beq\label{wkb}
\varphi(p_\eta, {\bf p}_\bot, \tau)\,=\,\frac{1}{\sqrt{2\omega( \tau)}}\,e^{\pm i\int \omega( \tau) d\tau}\,,
\eeq
which is a  solution to \eq{osc} in the WKB approximation. Using \eq{osc} we can verify that the WKB approximation $|\dot\omega|/\omega^2\ll 1$ is valid at all $\tau$ except at $\tau\to 0$. We will see below that the region $\tau\to 0$ does not contribute to the pair production. 

In the constant field ${\bf E} = E_z \hat{\bf z}$ we have according to \eq{ef} $\alpha=-E_z/2$. Assume for definiteness  that $\alpha<0$, i.\ e.\ the field points out to the positive $z$-direction. We have for the action
\beq\label{act}
S\,=\, \int \omega( \tau) d\tau\,=\,\int d\tau \sqrt{1/4\tau^2\,+\, p_\bot^2\,+\, \big(
p_\eta/\tau\,+\, e\,|\alpha|\, \tau\big)^2}\,.
\eeq
Changing the integration variable $\tau'=e|\alpha|\tau/p_\bot$ and introducing  a parameter $a=e|\alpha|/p_\bot^2$ we can cast the integral in Eq.~\eq{act} into form
\beq\label{act2}
S\,=\,\frac{1}{a}\int\frac{d\tau'}{\tau'}\sqrt{(\tau^{'2}+A_+^2)(\tau^{'2}+A_-^2)}\,,
\eeq
where 
\beq\label{apm}
A_\pm^2\,=\,\frac{1}{2}\,\bigg( 2p_\eta a\,+\,1\,\pm\,\sqrt{(2p_\eta a+1)^2\,-\,
a^2(1+4p_\eta^2)}\bigg)
\eeq

Consider now two cases depending on the sign of the determinant $D=(2p_\eta a+1)^2\,-\, a^2(1+4p_\eta^2)$. Case \emph{a}): $D>0$. In this case the integrand of \eq{act2} has a pole at $\tau'=0$ and two cuts along the imaginary axes $-A_+\le \im \tau' \le -A_-<0$ and $0<A_-\le \im \tau'\le A_+$.
In the WKB approximation the integration contour can be closed in the upper half plane \cite{LL3} in which case the imaginary part equals 
\begin{subequations}
\begin{eqnarray}
\im S_a&=&-\frac{1}{2a}\int_{A_-^2}^{A_+^2}\frac{d\tau_E^2}{\tau_E^2}
\sqrt{(A_+^2-\tau^2_E)(\tau_E^2-A_-^2)}\,=\,
-\frac{\pi}{4a}(A_+\,-\,A_-)^2\,\\
&=& -\frac{\pi}{4a}\bigg( 2p_\eta a+1-a \sqrt{1+4p_\eta^2}\bigg)
\label{ima}
\end{eqnarray}
\end{subequations}
where the Euclidean ``time" is $\tau_E=-i\tau'$.

Case \emph{b}) $D<0$.  $A_\pm$ become complex. It is convenient to 
change variables $iz=\tau'^2 +p_\eta a+\frac{1}{2}$ so that the action reads
\beq\label{actb}
S\,=\,- \frac{i}{2a}\int \frac{dz}{iz-(p_\eta a+\frac{1}{2})}\sqrt{(B-z)(B+z)}\,,\quad
B=i\sqrt{D}/2 \in\mathrm{Re}\mathcal{Z}\,.
\eeq 
The imaginary part of action arises from the discontinuity across the cut $-B\le \mathrm{Re}z\le B$ while the pole $z=-i(p_\eta a+1/2)$ does not contribute. We have
\begin{subequations}
\begin{eqnarray}
\im S_b&=&-\frac{1}{2a}\int dz \sqrt{(B-z)(B+z)}\left(\frac{i}{iz-(p_\eta a+\frac{1}{2})}\,-\,\frac{-i}{-iz-(p_\eta a+\frac{1}{2})} \right)\frac{1}{2i}\\
&=& -\frac{1}{2a}\int_{-B}^Bdz\frac{\sqrt{(B-z)(B+z)}}{z^2+(p_\eta a+\frac{1}{2})^2}
(p_\eta a+1/2)\\
&=&- \frac{\pi}{4a}\bigg(-2p_\eta a-1+a\sqrt{1+4p_\eta^2}\bigg)
\label{imb}
\end{eqnarray}
\end{subequations}
Combining the two cases \eq{ima} and \eq{imb} we derive the final result
\beq\label{ims}
\im S_+\,=\,-\,\frac{\pi}{4a}\big|\,2p_\eta a+1\,-\,a \sqrt{1+4p_\eta^2}\,\big|
\eeq

Eq.~\eq{ims} is not invariant under  the transformation $p_\eta\rightarrow -p_\eta$ which implies by means of \eq{fourier} that $\phi$ is not symmetric with respect to  pseudo-rapidity inversion. This is an artifact of our choice of the electric field direction: $\alpha<0$ means that ${\bf E}$ is pointing out in the direction of the positive $z$-axes (hence the subscript ``+" of $S$ in \eq{ims}). 
However, there is no preferred direction of the field since color charges are distributed approximately randomly in a nucleus. When the field points out to the negative $z$-direction the imaginary part of action reads:
\beq\label{ims2}
\im S_-\,=\,-\,\frac{\pi}{4a}\big|-2p_\eta a+1\,-\,a \sqrt{1+4p_\eta^2}\,\big|
\eeq

\begin{figure}[ht]
    \begin{center}
        \includegraphics[width=9cm]{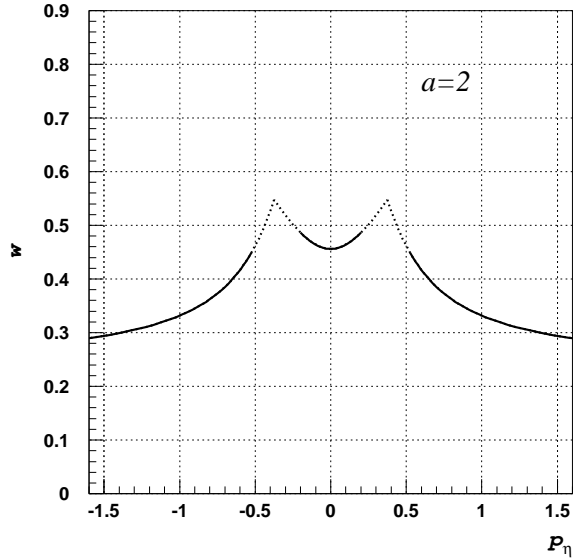}
\end{center}
\caption{Rate of pair production $w$ at $a=2$ as a function of $p_\eta$.}
\label{fig}
\end{figure}

The total probability of pair production is therefore
\beq\label{totprob}
w\,=\,\frac{1}{2}\left( e^{-2\im S_+}\,+\,e^{-2\im S_-}\right)\,.
\eeq
Note, that $2p_\eta a+1\,-\,a \sqrt{1+4p_\eta^2}> 0$ when $p_\eta> (a^2-1)/4a$ and negative otherwise, whereas $  -2p_\eta a+1\,-\,a \sqrt{1+4p_\eta^2}>0$ when $p_\eta<(1-a^2)/4a$ and negative otherwise. Therefore, we derive
\begin{subequations}\label{prob}
\beq\label{probp}
w\,=\,e^{-\pi|p_\eta|}\cosh\left(\frac{\pi}{2a}(1-a\sqrt{1+4p_\eta^2}\,)\right)\,,
\quad \mathrm{if}\,\,\, |p_\eta|>|a^2-1|/4a\,.
\eeq
\beq\label{probm}
w\,=\,\exp\left(-\frac{\pi}{2a}\big|1-a\sqrt{1+4p_\eta^2}\,\big|\right)\cosh(\pi p_\eta)\,,
\quad \mathrm{if}\,\,\, |p_\eta|<|a^2-1|/4a\,.
\eeq
\end{subequations}
The pair production probability as a function of $p_\eta$ is demonstrated in \fig{fig}.
%%%%
%%%%%

It is clearly seen from \fig{fig} that the derivative of $w$ with respect to $p_\eta$ is divergent at the cusps located at $p_\eta=\pm |a^2-1|/4a$. Therefore, the
quasi-classical approximation holds only in the three regions away from the cusps. 

\begin{figure}[ht]
    \begin{center}
        \includegraphics[width=10cm]{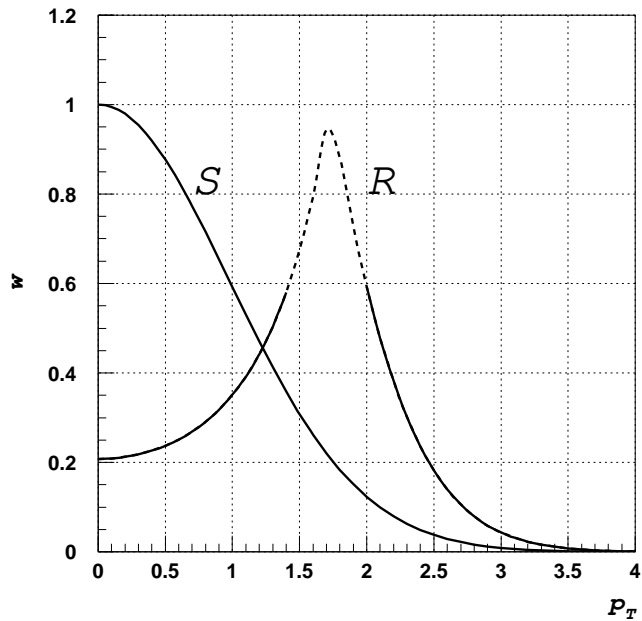}
\end{center}
\caption{Rate of pair production $w$ at $p_\eta =0$ as a function of $p_T$. The field strength is set to be $E= 6/e\,$GeV$^2$ so that $Q_s^2\simeq eE/\pi \approx 2$GeV$^2$. Line marked by `R' is the result is the Rindler-space calculation as given by \eq{prob}, whereas the line marked by `S' is the Schwinger formula. Dashed part of `R' corresponds to the values of $p_T$ at which the quasiclassical calculation breaks down. }
\label{fig:spectr}
\end{figure}

For massless particles $p_\eta=0$ and we have using \eq{probm}
\begin{subequations}\label{lim1}
\beq\label{lima}
w\,=\,e^{-\frac{\pi p_\bot^2}{eE_z}}\,,\quad p_\bot^2\gg eE_z\,,
\eeq
\beq\label{limb}
w\,=\,e^{-\frac{\pi}{2}}\,,\quad p_\bot^2\ll eE_z\,.
\eeq
\end{subequations}
The large $p_\bot$ tail of the probability distribution $w$, Eq.~\eq{limb}, coincides with the Schwinger formula. However, in general, the spectrum of produced pairs in comoving coordinates is significantly different from the Schwinger formula even for massless particles (with $p_\eta=0$), see \fig{fig:spectr}.

%%%%
%%%%%

%%%%%%%%%%%%%%%%%%%%%%%
\section{Non-Abelian case}

To generalize the obtained results for the case of $SU(2)$ gauge interaction we write down equations of motion for the gauge field $W_\mu$ in the external field
\beq\label{eqmot}
(-D^2\,\delta_{\mu\nu}\,+\, 2\,i\,e\, F_{\mu\nu})\,W_\mu\,=\,0\,,
\eeq
where the gauge condition $D_\mu A^\mu=0$ has been used. Here we made a separation of the $SU(2)$ gauge field  into two parts:  a quantized field $W_\mu$ and the classical background $A_\mu$. 

Eqs.~\eq{eqmot} can be diagonalized with respect to 
the light-cone gauge field components $W_\pm=W_0\pm iW_z$ 
\beq\label{diag}
-(D^2\,\pm\,2\,e\,E_z)\,W_\pm\,=\,0\,.
\eeq
Solving these equations in the Rindler coordinates yields for $W_\pm$ the same equation as for $\phi$, see \eq{fourier},\eq{osc} but with the shifted frequency $\omega^2\rightarrow\omega_\pm^2=\omega^2\pm 2eE_z$. Therefore, the pair-production probability in the case of $SU(2)$ can be read off the Eqs.~\eq{probp} and \eq{probm} with $a$ replaced by $a_\pm$, where
\beq\label{apmw}
a_\pm\,=\,\frac{eE_z}{2\,(p_\bot^2\pm 2\,eE_z)}\,,
\eeq
and sum over polarizations is performed in the arguments of $\cosh$ and $\exp$:
\begin{subequations}\label{probsu2}
\beq\label{probpsu2}
w\,=\,e^{-2\pi|p_\eta|}\cosh\left(\sum_\pm\frac{\pi}{2a_\pm}(1-a_\pm\sqrt{1+4p_\eta^2}\,)\right)\,,
\quad \mathrm{if}\,\,\, |p_\eta|>|a_\pm^2-1|/4a_\pm\,.
\eeq
\beq\label{probmsu2}
w\,=\,\exp\left(-\sum_\pm\left| \frac{\pi}{2a_\pm}\big(1-a_\pm\sqrt{1+4p_\eta^2}\,\big)\right|\right)\cosh(2\pi p_\eta)\,,
\quad \mathrm{if}\,\,\, |p_\eta|<|a_\pm^2-1|/4a_\pm\,.
\eeq
\end{subequations}
We see that  in non-Abelian case $0\le a_\pm\le 1/4$.
As the result, asymptotic behavior of the pair-production probabilities is slightly modified with respect to the Abelian case \eq{lima},\eq{limb} (for $p_\eta=0$)
\begin{subequations}\label{lim2}
\beq\label{lima2}
w\,=\,e^{-\frac{2\pi p_\bot^2}{eE_z}}\,,\quad p_\bot^2\gg eE_z\,,
\eeq
\beq\label{limb2}
w\,=\,e^{-4\pi}\,,\quad p_\bot^2\ll eE_z\,.
\eeq
\end{subequations}
Large values of $\im S$ insure applicability of the quasi-classical approximation in these asymptotic regions.

%%%%%%%%%%%%%%%%%
\section{Effect of back-reaction}
\label{sec:br}

To estimate the back-raection of produced particles on the background field note that the only non-vanishing light-cone components of the field-strength tensor are $F^{+-}=-F^{-+}=E_z$, where 
$E_z=-\frac{1}{\tau}\partial_\tau(\alpha(\tau)\tau^2)$. Therefore, Eqs.~\eq{eqmot2} can be written as
\beq\label{rew}
-\frac{\partial E_z}{\partial x_-}\,=\,j^+\,,\quad 
\frac{\partial E_z}{\partial x_+}\,=\, j^-\,.
\eeq
Switching  to the Rindler coordinates $\tau$ and $\eta$ where  pseudo-rapidity $\eta=\frac{1}{2}\ln\frac{x_+}{x_-}$ we find  from \eq{rew} that 
\beq\label{curr}
j_+\,=\, x_+j(\tau)\,,\quad j_-\,=-\, x_-j(\tau)\,,
\eeq
which  satisfies the current conservation $\partial_\mu j^\mu =0$ for arbitrary $j(\tau)$, provided that $j_\mu$ is independent of the transverse coordinates. 
Finally, we obtain
\beq\label{br}
\partial_\tau E_z\,=\, -j(\tau)\tau\,.
\eeq
The back-reaction effect can be neglected if $ |\partial_\tau E_z|\,\ll |E_z|/\tau$ which implies that 
\beq\label{qc}
|j(\tau)|\,\ll\, \frac{1}{\tau^2}\,|E_z|\,.
\eeq
This is equivalent to requiring that 
the energy of interaction of the current with the external field $j^\mu A_\mu$ be much smaller than the energy of the field $E_z^2/2$ itself.

To check the validity of the adiabatic approximation \eq{qc} we need to calculate the current induced by the external field. Substituting \eq{fourier} and \eq{wkb} into \eq{tok}  we derive  in a classical limit
\beq\label{meancur}
\langle j(\tau)\rangle\,=\,\frac{1}{\tau^2}\,g\,\int\frac{dp_\eta\, d^2{\bf p}_\bot}{(2\pi)^3
\,\omega(\tau)}\,\bigg( \frac{p_\eta}{\tau}\,-\, e\,\alpha\,\tau\bigg)\, 2w(p_\eta,{\bf p}_\bot)\,,
\eeq
where $g$ is the number of degrees of freedom of each of the produced particles. Formally, integral over $p_\eta$ diverges (since $w$ is constant at large $p_\eta$). However, the rhs of \eq{meancur} with $w=1$ is just a vacuum term. It vanishes by reflection symmetry for the fixed integration boundaries of the kinetic momentum $p_\eta-e\alpha\tau^2$ \cite{Kluger:1991ib}.

In the physical situation we are interesting in (gluon production) $p_\eta\approx 0$. Thus, we use Eqs.~\eq{lim2} to estimate the induced current \eq{meancur}.
At later times $\tau\gg p_\bot^{-1}, (eE_z)^{-1}$ integration over $|p_\eta|\le\frac{1}{\pi}$ and $p_\bot^2\in[0,\infty)$ in \eq{meancur} yields
\beq\label{indcur}
\langle j(\tau)\rangle \approx \frac{1}{\tau^2}\frac{2e E_zg}{(2\pi)^3}\, e^{-\frac{\pi}{2}}\,.
\eeq
Then, employing \eq{qc} we conclude that the back-reaction effect is small
if the following condition is satisfied
\beq\label{cond}
e\,g\,\ll\,4\pi^3 e^{\frac{\pi}{2}}\,.
\eeq
In the non-Abelian case the similar calculation results in even weaker condition
\beq\label{conda}
e\,g\,\ll\,8\pi^3 e^{4\pi}\,,
\eeq
which is satisfied for all reasonable values of coupling $e$ and number of degrees of freedom $g=2(N^2-1)$. Note, that the average square of the transverse momentum of the produced particles is given by $\langle p_\bot^2\rangle=eE_z/2\pi$, see \eq{limb2}. Therefore, the estimate  \eq{conda} is valid at  $\tau\gg 1/\sqrt{2\pi}\langle p_\bot \rangle$, i.e. at all times of interest in heavy-ion collisions.

%%%%%%%%%%%%%%%%%%%%%
\section{Discussion}\label{sec:disc}

The central result of our paper is Eqs.~\eq{prob} and \eq{probsu2}. They solve the problem of particle production by a constant electric field in the Rindler coordinates in absence of back-reaction of produced pairs on the field. 
We observed that the pair production probability in Rindler coordinates is essentially different from that observed by an inertial observer as given by the Schwinger formula, \fig{fig:spectr}. Although the classical background field is boost invariant, the quantum corrections strongly violate it. 

It is important to emphasize, in order to avoid a possible confusion, that had we chosen to quantize the dynamical fields $\phi$  and $W_\mu$ in Minkowski coordinates and use the quasi-classical approach discussed in this paper, we would have obtained Schwinger's formula (see e.\ g.\  \cite{Dunne:2004nc}). 
In the present paper, the dynamical fields are assumed to be quantized in the Rindler coordinates $\tau$ and $\eta$. At any given Rindler time $\tau$, the quasi-classical trajectories are $x_+x_-=\mathrm{const.}$ corresponding to the uniform acceleration in Minkowski space. On the contrary, Minkowski observers move along the straight  lines.  As a result, an observer in a Rindler frame is not equivalent to an observer in a Minkowski frame leading, in particular,  to the famous Unruh effect \cite{Unruh:1976db}.  In the present paper we argued that the pair production probability in a constant background field in Rindler space is also different from that in Minkowski space.

The pair production probability $w$ is proportional to the particle phase space density and can be used as an initial condition for the subsequent hydrodynamical evolution. It has been argued in \cite{Romatschke:2005pm} that small deviations from the boost invariance are amplified by the Weibel instability and can speed up the process of thermalization in high energy heavy-ion collisions.
Therefore, our results can be important for understanding the final spectra of particles produced in heavy-ion collisions.

Our estimate of back-reaction shows that the induced current is small. However, the quasi-classical approach to the back reaction is admittedly approximate. It ignores both  the produced pairs correlations which may induce the background field oscillations \cite{Kluger:1991ib} and quantum fluctuations which has to be properly renormalized \cite{Cooper:1989kf}. We are going to address these important issues in forthcoming publications. 

It is also important to generalize the obtained results to include  fermions. This will allow us to address the question of heavy quark production at small $p_\bot$ and shed light on the heavy quark multiplicities.

%%%%%%%%%%%%%%%%%%%%%%%%%%%%%%%%%%%%%%%%

\vskip0.3cm
{\bf Acknowledgments}
%\vskip0.3cm
The author is grateful  to Lev Kofman, Dima Kharzeev and Emil Mottola  for informative discussions of related problems.  The author would like to thank RIKEN, BNL and the U.S. Department of Energy (Contract
No. DE-AC02-98CH10886) for providing the facilities essential for the
completion of this work.

%%%%%%%%%%%%%%%%%%%%%%%%%%%%%%%%%%%%%%%%%%%%%%%%%%%%%%%%%%%%%%%%%%%%%%%%%%%%


\begin{thebibliography}{99}
\bibitem{com} The transverse momentum dependence of the background field is crucial for understanding the interplay between the pair production mechanism and the perturbative  particle production. Unfortunately, no method is known which would allow to derive the pair production in a background field depending both on time and coordinate(s).
\end{thebibliography}
\end{document}